
%
%
%

\voffset=6truemm 
\hoffset=35truemm   
\hsize=134truemm   
\vsize=212truemm   
\pretolerance=5000   
\raggedbottom     
\raggedright     
\baselineskip=12.045truept  
\parindent=2.5em        
%
\def\apj{ApJ}

\def\mnras{MNRAS}
\def\aap{A\&A}
\def\aj{AJ}

\def\kms{{\rm\,km\,s^{-1}}}

\def\eg{{\it eg.,\ }}

\def\pp{\parshape 2 0truecm 13.4truecm 1truecm 12.4truecm}
%
\def\apjref#1;#2;#3;#4 {\par\pp#1, {#2}, {#3}, #4 \par}

\def\ltsima{$\; \buildrel < \over \sim \;$}
\def\simlt{\lower.5ex\hbox{\ltsima}}
\def\gtsima{$\; \buildrel > \over \sim \;$}
\def\simgt{\lower.5ex\hbox{\gtsima}}
\def\ub{\underbar}

\def\ls{\vskip 12.045pt}   
\def\ni{\noindent}        
\def\et{{\it et\thinspace al.}\ }    
\def\etal{\et}

\def\deg{\ifmmode^\circ\else$^\circ$\fi}    

\def\arcs{\ifmmode {'' }\else $'' $\fi}     
\def\arcm{\ifmmode {' }\else $' $\fi}     
\def\buildrel#1\over#2{\mathrel{\mathop{\null#2}\limits^{#1}}}
\def\mper{\ifmmode \buildrel m\over . \else $\buildrel m\over .$\fi}
\def\hper{\ifmmode \rlap.^{h}\else $\rlap{.}^h$\fi}
\def\sper{\ifmmode \rlap.^{s}\else $\rlap{.}^s$\fi}
\def\arcsper{\ifmmode \rlap.{' }\else $\rlap{.}' $\fi}
\def\arcmper{\ifmmode \rlap.{'' }\else $\rlap{.}'' $\fi}


\vglue 29.10pt 
\def\section#1{\ls\ni\ub{#1}\ls}
\def\subsection#1{\ls\ni\ub{#1}}
\item\item{THE FUTURE OF GALAXY
POPULATIONS\footnote{$^\heartsuit$}{Proc.~`Clusters, Lensing and the
Future of the Universe', ed.~V.~Trimble (ASP) in press.} \hfil}

\ls\ls

ROSEMARY F.~G.~WYSE

\item\item{{Department of Physics and Astronomy,
Johns Hopkins University,  \break \hfil}
Baltimore, MD 21218 USA (permanent address)}

\item\item{{Center for Particle Astrophysics, Le Conte Hall,
\break \hfil}
UC Berkeley, CA 94720 }

\ls\ls
\item\item{\ub{ABSTRACT}\hskip 5mm What the
future holds for the stellar populations of the Milky Way Galaxy may
(optimistically) be
predicted by study of their past histories, as written in the chemical
abundance distributions, angular momentum distributions, velocity dispersion
tensor, ages and spatial structure of their constituent older stars.
}

\ls\ls\ls\ls

\ni\ub{INTRODUCTION}

\ls
\ni
The study of the spatial distribution, kinematics and chemical
abundances of stars in the Milky Way Galaxy constrains models of
disk galaxy formation and evolution.  One can address
specific questions such as
\item {(i)} When was the Galaxy assembled?  Is this an ongoing
process? What was the merging history of the Milky Way?
\item {(ii)} When did star formation occur in what is now `The
Milky Way Galaxy'?  Where did the star formation occur then?
What was the stellar Initial Mass Function?
\item {(iii)} What are the relationships among  the different
stellar components of the Galaxy?

The nature of Dark Matter determines the way in which structure forms in the
Universe.  The popular theory of Cold Dark Matter (CDM; \eg\ reviewed by
Silk and Wyse 1993) predicts that the first objects to collapse under
self-gravity are a small fraction of the mass of a typical galaxy, so that
galaxies form by clustering and merging of these smaller objects.  All
density fluctuations are initially just gas and dark stuff. The star
formation histories in these `building-blocks' can be very varied, and at
present cannot be predicted with any level of certainty by theory. However,
feedback from stars plays a major role in the energy balance.
The rate of merging and growth of mass of a
protogalaxy can be estimated by studying the dark matter, which
is assumed to be dissipationless and hence to have simpler physics than the
baryonic component. Analytic calculations of the rate at which
structure grows in dark halos agree remarkaby well with that seen in N-body
simulations (\eg\ Carlberg 1990; Lacey and Cole 1993, 1995).  Typically at a
redshift of unity, a galaxy would be $\simlt 2/3$ of its present mass
(Carlberg 1995).
Consideration of dark matter haloes
most probably underestimates the
longevity of individual baryonic structures which would be
observed as distinct galaxies, since they can dissipate and thereby reduce
their merging cross-section.
Summers (this volume) summarizes the complex
physics that one may model with sophisticated hydrodynamic codes.

Lacey and Cole (1993) have a particularly vivid schematic
representation of the merging process as a tree (their Figure 6),
where time increases from top to bottom, and the width of the branches
indicates the mass of a particular halo associated with galactic
substructure.  The merging history of a galaxy is then described by
the shape of the tree. The extremes of morphological type may perhaps
be the result of merging histories that are described by the two types
of trees that I at least was taught to draw as a child -- either one
main trunk from top to bottom, with many small branches joining the
trunk at all heights, or a main trunk that splits into two repeatedly.
The latter, dominated by `major mergers' or equal mass mergers, may
lead to an elliptical galaxy.  The former, where the merging history
is dominated by `minor mergers', or very unequal mass mergers with a
well-defined central core at all times, may lead to a disk galaxy.
This picture of disk galaxy formation -- building up by accretion of
substructure onto a central core -- provides a synthesis of elements
of the much-discussed and previously apparently mutually exclusive
`monolithic collapse' paradigm of Eggen, Lynden-Bell and Sandage
(1962) and the `chaotic' halo formation envisaged by Searle and Zinn
(1978).

I will discuss recent observations of the stellar populations of the Milky
Way with an aim to understanding their past evolution, as a means to
predicting their future.

\ls
\ls
\ni\ub{THE STELLAR HALO}
\nobreak
\ls
\ni

 The hierarchical clustering and merging picture of galaxy formation
predicts that there should be many shredded satellite galaxies for
every parent galaxy. Indeed, the field stellar halo would
consist of disrupted smaller-scale structure. Observational evidence
has been accumulating that merging of smaller systems has played a
role in the evolution of the stellar halo.  The discovery of the
Sagittarius dwarf spheroidal galaxy (Ibata, Gilmore and Irwin 1994),
apparently in the process of being digested by the Milky Way, argues
quite irrefutably for on-going accretion. Stars which become unbound
from the outer regions of satellite galaxies, beyond their tidal
radius set by the Milky Way gravitational potential, will remain on
orbits close to that of the satellite galaxy at the time of their
evaporation.  Thus, provided that the satellite is tidally disrupted
prior to significant orbital decay through dynamical friction,
shredded stars may be expected to contribute to the stellar halo.  A
typical dwarf spheroidal galaxy contains a significant fraction of its
stars in intermediate-age or young populations, despite a low
metallicity.  For example, Smecker-Hane {\it et al.} (1994)
demonstrated definitively that the majority of stars in the Carina
dSph were of age $\sim 8$ Gyr, and Lee {\it et al.} (1993) have shown
that the age of the dominant population in Leo I is rather young, at
$\sim$ 3 Gyr. Few, if any, have exclusively old populations. This
intermediate-age, metal-poor population provides a distinctive
signature of their contribution to the field halo.

\vfill\eject
\subsection{The Age(s) of the Stellar Halo}

\ni
One of the defining characteristics of the field population II is its
clear lack of massive main sequence stars (Sandage 1969 and refs
therein).  The distinct turn-off of the stellar halo, at B$-$V$\simeq
0.4$, is a striking feature of deep star counts (e.g. Gilmore and Wyse
1987, their Fig. 3).  This turn-off color corresponds to an age,
derived from comparison with stellar isochrones, of $\simgt 15$Gyr,
for a population with the mean metallicity of the stellar halo,
$<$[Fe/H]$> \sim -1.6$ dex. Any halo stars significantly younger than
this are restricted to a tracer population.  A few high-velocity dwarf
stars bluer than the halo turn-off were identified in early surveys
(e.g. Bond and MacConnell 1971; the few high-latitude B main-sequence
stars identified by Conlon {\it et al.} ( 1992) are metal rich and
apparently unrelated to the field halo population).  More recently,
Preston, Beers and Schectman (1994) concluded that a significant
fraction of their sample of stars, selected on the basis of the
weakness of their Calcium H and K lines, were dwarf stars bluer than
the halo turnoff.  The absolute normalisation is difficult to quantify
given the selection criteria of the sample, but they suggest that
perhaps 4\%--10\% of the stellar halo is in this component, which they
attribute to the accretion of dwarf spheroidal satellites. Evidence in
favor of accretion also comes from the kinematic `moving groups' which
have been tentatively identified in the Galactic halo (e.g. Arnold and
Gilmore 1992; Majewski 1993).

Quantification of the fraction of the stellar halo that is younger than
the dominant population requires careful analysis of samples where the
selection effects are understood.
The recent survey by Carney {\it et al.} (1994 and references therein)
contains the largest sample of halo stars (1447 stars), selected on the
basis of proper motions, for which accurate abundances and photometry are
available.  Adopting a working definition of the metal-poor halo for the
moment as stars in this sample with  [Fe/H] $\simlt -1$, yields  477 stars.
The relationship between $B-V$ color and [Fe/H] for this sub-sample is
indicated in Figure 1, taken from Unavane, Wyse and Gilmore (1995). The
color distribution in this figure  shows clearly the turnoff of an old,
metal-poor population, with the turnoff color being a function of [Fe/H],
just as the isochrones behave. Main sequence stars which are bluer than a
given isochrone (at fixed metallicity) are younger than the age of that
isochrone.  There is clearly room for a couple of Gyr scatter about the age
of the isochrone defining the dominant population, especially since the
B$-$V turn-off colors of the old isochrones crowd together. However, the
$\simgt 15$ Gyr isochrones adequately represent the cut-off color for the
majority of the stars; bluer stars are rare.

One may determine a firm {\it upper limit\/} to
the number of field halo stars that are much younger than the dominant
old population by counting all stars that lie to the blue of a chosen
fiducial, old isochrone, taking account of the observational
uncertainties to assign a probability  to each star that it is bluer,
and then weighting this number by the luminosity function to include
the lower mass, redder stars of the younger population. Counting {\it
all\/} stars that are bluer includes, e.g., blue stragglers of the
type found in globular clusters etc. This exercise (see Unavane {\it
et al.}) leads to the conclusion that the upper limits on younger star
population in the field halo comprises $\sim 3$\% in the most
metal-poor range, [Fe/H] $ < -1.95$; $\sim 6$\% in the middle range,
and $\sim 28$\% in the most metal-rich range considered, $-1.5 \le $
[Fe/H] $ < -1$.

\ls\ls\ls\ls
\ni
This is a sample selected by high proper-motion, and although as
mentioned above one may expect stars to be removed from low density
satellite galaxies prior to much orbital decay, here is no robust {\it
a priori\/} expectation that the kinematics of any young halo
population should be closely similar to those of the older halo stars.
However, Ryan and Norris (1993) investigated the probability that a
star with a given space motion would pass into a proper-motion
selected sample of given selection criteria.  In situations where the
lag in rotational velocity of the parent population behind that of the
Sun dominates, the survival probabilities, albeit low, do not vary by
more than a factor of a few, over the range $v_{lag} \sim 100 - 200$
km/s.

\subsection{How Many dSph Could Have Been Accreted?}

\ni
Unavane {\it et al.} discuss how this derived fraction of young stars
in the halo may then be re-expressed in terms of the number of
satellite galaxies that could have been accreted, given the
characterization of the stellar populations of the satellites.  A
direct comparison of the color distribution of the Carney {\it et al.}
sample with the appropriate metallicity ($-1.95 \le$ [Fe/H] $ <
-1.55$, equal to the range in the Carina dwarf) and of the stars in
the Carina dwarf may be made, circumventing model and isochrone
dependencies. They concluded that 3\% of these halo stars are
consistent with being drawn from a parent population similar to that
of the Carina dwarf; renormalising to the entire halo metallicity
distribution reduces this to around 1\%.
The Carina dSph has a total luminosity of $M_V \simeq -9.2$, while the
halo has $M_V \simeq -17.5$.  Thus this limit of 1\% of the halo
amounts to $\simgt 20$ Carina dwarfs.

Extending this result to constrain the recent accretion of {\it all\/}
dwarf galaxies drawn from the present population of satellite galaxies
would require detailed color and metallicity data for each of the
satelite galaxies.  Detailed deep photometric data are not in general
available, but reliable metallicity estimates are.
Figure 2 shows the halo metallicity distribution function taken from
Laird {\it et al.} (1988), together with the mean metallicities of the
seven dSph companions to the Milky Way for which reasonably reliable
spectroscopic metallicity estimates are available. The composite
stellar population of dSph complicates the determination of
photometric metallicity estimates from analysis of their
color-magnitude diagrams.

\ls\ls\ni
The upper panel shows these distributions simply normalised to unity
in the highest bin, with  the Magellanic Clouds  as
indicated. The dwarf companion galaxies to the Milky Way  have
an approximately uniform distribution of metallicity (and indeed of
luminosity, unlike the simplest predictions from CDM cosmology which
predict a steep mass function in this range).  The lower panel
compares the {\it luminosity-weighted\/} metallicity distribution for
the dSphs with the halo metallicity distribution. It is this weighted
distribution function which would correspond to the metallicity
distribution of stars accreted into the halo from a random subset of
the present day dSph luminosity function. It clearly does {\it not}
match the metallicity distribution function of the field halo. While
this weighting is rather uncertain, as all dSph parameters are still
only approximately known, the luminosity--metallicity relation (e.g.
Armandroff {\it et al.} 1993) ensures that Fornax and Sagittarius
dominate, providing a relatively metal-rich mean to the stars that
make up the present dSph population.  This weighting
excludes the Magellanic Clouds, which are completely inconsistent with
providing any contribution to the field halo.

 The signature of recent accretion from the present parent population of
dSph would be intermediate-age stars in the halo, with metallicity strongly
peaked at [Fe/H] $\simgt -1.5$ dex. This is indeed the metallicity
distribution for the candidate young halo stars, in Figure 1.  Detailed age
distributions are lacking for the metal-rich dSph, but one can derive a
rough limit on the number of such that could have been assimilated into the
halo, by assuming 100\% intermediate-age.  The $\sim 28$\% contribution of
the stars in the Carney {\it et al.} sample in the metallicity range $-1.55
\le $ [Fe/H] $ < -1$ younger than $\sim 15$Gyr corresponds to approximately
10\% of the entire halo (using the Laird {\it et al.} halo metallicity
distribution) being `young'. The field halo is approximately 50 times more
luminous than Fornax, adopting a luminosity of $-13.2$ for Fornax, so that
of order 5 systems similar to Fornax are implicated.

\subsection{Other Signatures of dSph Accretion}

\ni
The range of ages and chemical abundances within these systems
indicate a level of self-enrichment, which implies some retention
and/or re-capture of gas (e.g. Silk, Wyse and Shields 1987).  As
discussed in Unavane {\it et al.} (1995), the star formation history
of a typical dSph may impart an observable signature in the elemental
abundance ratios, specifically evidence of enrichment by Type Ia
supernovae.  This is to be contrasted with the three-times-solar value
of the oxygen-to-iron ratio seen in halo field stars, generally
attributed to enrichment by Type II supernovae alone (e.g. Wyse and
Gilmore 1988; Nissen {\it et al.} 1994).
Specific observations
of the element ratios in candidate `young' halo stars have not been
made with modern detectors, to be analysed with up-to-date stellar
atmospheres.  This will be an important observation.

It is impossible at present to say anything reliable from carbon
star statistics due to lack of data, particularly for the field halo.

\subsection{The Future of the Stellar Halo}

\ni
The past of the stellar halo apparently involved assimilation of stars
that plausibly were formed in satellite galaxies, albeit not amounting
to a large fraction of the luminous mass of the halo.  Such accretion
of dSph is apparently on-going, and may be expected to continue into
the future.  Thus one may conclude that the future of the field
stellar halo is to get (relatively) younger.

\ls
\ls
\ni\ub{THE CENTRAL BULGE}
\nobreak
\ls
\ni
The central regions of galaxies are obvious potential
repositories of accreted systems, being the bottom of the local
potential well, provided the accreted systems are sufficiently
dense to survive disruption while sinking to the center (e.g.
Tremaine, Ostriker and Spitzer  1975).
The mean metallicity of the
bulge is now reasonably well-established at $<$[Fe/H]$> \sim -0.3$ dex
(McWilliam and Rich 1994; Ibata and Gilmore 1995), with a
significant spread to below $-1$ dex, and to above solar.  Thus
satellite galaxies that could have contributed significantly to the bulge
are restricted to those of high metallicity, more like the Magellanic
Clouds, or the most luminous dSph.

The Sagittarius dwarf spheroidal galaxy was discovered through
spectroscopy of a sample of stars selected purely on the basis of
color and magnitude to contain predominantly K giants in the Galactic
bulge.  After rejection of foreground dwarf stars, the radial
velocities isolated the Sagittarius dwarf galaxy member stars from the
foreground bulge giants.  Not only the radial velocities distinguish
the dwarf galaxy, but also its stellar population -- as seen in Figure 1
of Ibata {\it et al.} (1994), all giant stars redder than B$_J -$R$\simgt
2.25$ have kinematics that place them in the low velocity-dispersion
component {\it i.e.} in the Sagittarius dwarf. This is a real
quantifiable difference between the {\it bulge} field population and
this dwarf spheroidal galaxy, and demonstrates that the bulge could
not have formed by accretion of such systems.

As with the halo above, the carbon star population of the bulge
can be compared with those of typical extant satellites.  In this
case there is a clear discrepancy between the bulge and the
Magellanic Clouds and the dSph (Azzopardi and Lequeux 1992).

The observation that the mean iron abundance of the stellar halo is
well below the theoretical yield for a solar neighborhood IMF is most
easily explained by gaseous outflow, truncating chemical evolution
(Hartwick 1976).  The stellar bulge could then form from this gas.
Wyse and Gilmore (1992) demonstrated that the specific angular
momentum distributions of the stellar halo and bulge are extremely
similar, strengthening their association.  Further, the mass ratio of
stellar halo and bulge is as expected if the bulge formed from gas
removed from the halo (Carney 1990).  Elemental abundance data for
stars in the bulge, over the range of iron abundances seen, would
provide a good discriminant of the star formation history of both the
halo and bulge, in this picture (see Wyse and Gilmore 1992).  Such
data are becoming available (McWilliam and Rich 1994) although at
present the uncertainties are too large to make definitive statements.

An alternative, dissipationless formation picture for the central
bulge has gained favor recently, largely due to the images from the
{\sl COBE} satellite in which the bulge has a pronounced `peanut'
shape, and the analyses of these and other data, including gas
kinematics, that imply that the bulge is triaxial (Blitz and Spergel
1991; Binney {\it et al.} 1991).  This picture appeals to a buckling
instability of an initially-thin stellar disk (Combes {\it et al.}
1990; Raha {\it et al.} 1991).  The disk first forms a bar in its
plane, and the bar subsequently thickens through an out-of-plane
bending instability.  However, the latest analysis of the DIRBE data
(Weiland \etal 1994) concluded that the peanut morphology is an
artefact of varying extinction.  Indeed, all physical parameters of
the central regions of the galaxy remain rather uncertain; the bar
models in the literature do not successfully explain all of the
features in the gas kinematics (cf. Lizst 1995), while the inner disk
structure is clearly ill-defined. The scheduled 15$\mu$ survey of the
inner Galaxy with the {\sl ISO\/} satellite (PI A.~Omont) should
mitigate these circumstances.  Extant models for the bar provide
gravitational torques to the surrounding disk gas that can drive gas
inflows of amplitude $\simlt 0.1 M_\odot$/yr (Gerhard and Binney
1993).  This could fuel continuing star formation, providing a young
tail to the age distribution in the bulge, and probably stars with low
oxygen-to-iron.  The age distribution of stars in the bulge is
unfortunately rather uncertain at present, but is probably quite broad,
with a mean of $\simlt 10$~Gyr (e.g. Holtzman {\it et al.} 1993).

A lack of a chemical abundance gradient argues against dissipational
formation, and for such a dissipationless mechanism (the opposite is
not necessarily true).  Hence, the recent revision downwards of the
mean metallicity of the bulge, to a value that is remarkably similar
to that of K giants in the solar neighborhood (McWilliam 1990), and
indeed across the disk (Lewis and Freeman 1989) supports this latter
disk-instability mechanism. Further, Ibata and Gilmore (1995) find
that even within the bulge itself the data are consistent with no
gradient in mean metallicity out to 3kpc.  In this picture for bulge
formation,the similarity in specific angular momentum distributions of
halo and bulge, and their difference compared to those of the thick
and thin disks, would then have to be a coincidence.

It is extremely difficult to cast a horoscope for the bulge, other than to
say that as the bottom of the local potential well, it is likely to grow in
mass.

\ls
\ls
\ni\ub{THE THICK DISK}
\nobreak
\ls
\ni

The characteristics of the thick disk at the solar neighborhood
are reasonably well defined, but the global properties of this
stellar population are not.
Locally it can
be described by a predominantly  old population (Gilmore and Wyse
1987; Carney \etal\ 1990; Schuster and Nissen 1989; Gilmore, Wyse and Jones
1995), with mean
metallicity about one-third that of the Sun, and a significant
dispersion in metallicity, of $\sim 0.3 $ dex (Gilmore and Wyse
1985; Laird \etal 1988; Friel 1987; Morrison, Flynn and Freeman
1990; Gilmore, Wyse and Jones 1995).
It is a kinematically `warm' component, with vertical
velocity dispersion of around $45 {\rm km
s^{-1}}$, and
with a constant lag behind circular velocity of amplitude $\sim 50 {\rm km
s^{-1}}$, out to $z = 2.5 {\rm kpc}$ (Soubiran 1993).  The scaleheight of
the thick disk is a factor of $3-4$ larger than that of the old thin disk,
reflecting its higher velocity dispersion. Note although there is indeed an
age--velocity dispersion relationship for stars in the thin disk, the value
of the vertical velocity dispersion  saturates at $\sim 20 \kms$ for stars
older than a few Gyr (\eg Freeman 1991). This  may be understood if the
heating mechanism responsible is confined to the thin disk itself, such as
scattering by giant molecular clouds (\eg Lacey 1991).

  The thick disk of the Milky Way Galaxy at least
morphologically could be a minor-merger remnant; provided all the
orbital energy of an accreted satellite, mass $M_{sat}$,
is available to increase
the random energies of the stars in a thin disk of mass
$M_{disk}$, then after a merger the thin disk will be heated by
an amount
(Ostriker 1990) $$\Delta v_{random}^2 = v_{orbit}^2 M_{sat}/M_{disk}.$$
Of course  the internal degrees of freedom of
the satellite could also be excited and any gas present could,
after being heated,
cool by radiation,  so this is a definite upper
limit to the heating of the disk.  This estimate is suggestive, however,
that the thick disk,
which has a vertical velocity dispersion of $\sim 45 \kms$, could
be formed from a thin disk with vertical dispersion of $\sim 20
\kms$ by accretion of a satellite of about 10\% of the mass of
the disk.  T\'oth and Ostriker (1992) extended these ideas and argued
that the observed thinness of the stellar disk of our Galaxy could be
used to limit the allowable accretion of stellar satellites since the
birth of the Sun to only a few percent of the mass of the disk.  They
then argued that this low accretion rate favored a low density CDM
Universe, since the growth of perturbations would then be truncated at
a redshift of $\sim \Omega^{-1}$.

  Indeed, Quinn, Hernquist and Fullagher (1993) showed through N-body
simulations that the accretion of a 10\% by mass satellite,
sufficiently robust to survive many passages through the disk, could
produce a thick disk that had several similarities to that of the
Milky Way.  This thick disk consists of both heated formerly thin-disk
stars and shredded-satellite stars; the mix of these obviously depends
on the (many) model parameters, both orbital and internal to the
galaxies.  The sensitivity of the end-point of a merger to the input
was emphasised by Carlberg (1995), who demonstrated that a less dense,
arguably more realistic, satellite than the rather extreme case
(density within the half mass radius of the satellite fully 75\% of
that of the inner disk) considered by Quinn {\it et al.} could impart
essentially no heating.  This sensitivity makes constraining $\Omega$
from disk thickness rather dangerous.  However, the large dispersion
in the properties of thick disks identified in external galaxies (e.g.
van der Kruit and Searle 1982; Shaw and Gilmore 1988; Morrison,
Boroson and Harding 1994) is perhaps most readily understandable in
the context of a merging scenario, since so many outcomes are possible
in this case, as opposed to other scenarios, such as a cooling
instability (e.g. Wyse and Gilmore 1988).

The Large Magellanic Cloud  is perhaps the type of satellite modelled by
Quinn {\it et al.}  The Magellanic Clouds are indeed most probably
interacting tidally with the Milky Way;
the Small Magellanic Cloud appears to be significantly
distended along the line of sight,
and the Magellanic Stream is plausibly also a tidal
effect.  Proper motion estimates for the LMC (Kroupa {\it et al.} 1994;
Jones {\it et al.} 1994) suggest even more significant
interaction in the future.   Indeed, Lin {\it et al.} (1995) predict a
peri-Galacticon of less than 10kpc for the LMC, about 8Gyr in the future.
This will have a profound effect on the present thin disk.

Thus one can say with reasonable confidence that the future of the thick
disk is to get more massive, as the now-thin disk is heated.

\ls\ls
\ni\ub{THE THIN DISK}

\ls
\ni
Many of the major observational advances in recent years in the
investigation of the evolution of the thin disk of the Milky Way have
come from the analysis by Edvardsson {\it et al.} (1993) of the ages,
kinematics and elemental abundances of their sample of F/G dwarf
stars, observed in the solar neighborhood.  The observational errors
are small enough that one can state with confidence that there is
intrinsic scatter in the iron-abundance--age relation; indeed, over
much of the lifetime of the disk, the scatter overwhelms any mean
trend.  The scatter remains even if one isolates a subsample likely to
have been formed in a narrow range of Galactocentric distance. Star
formation and chemical enrichment evidently proceeded in a rather
inhomogeneous manner.

In keeping with the cosmological leaning of this conference, Figure 3
shows the age--iron abundance data for the Edvardsson {\it et al.}
subsample which they identify as likely to have formed close to the
solar circle, but with stellar age transformed to redshift of
formation.  Their age estimates are sufficiently old that I have
taken a Hubble constant of H$_o = 42$~km/s/Mpc, with $\Omega = 1$.
The  errors in (log) age, of 0.1 dex, translate into large
uncertainties at redshifts greater than about unity; I have suppressed the
oldest stars in the plot.  The
main point is clear, however, which is that Galactic astronomers are
not surprised that there are metals in absorber systems along the line
of sight to distant quasars.

 The mean global star-formation rates in disk galaxies have decreased
only slowly over the last Hubble time, and it is likely that they can
sustain their present star formation for the forseeable future, over
many Gyr (e.g. Kennicutt,  Tamblyn and Congdon 1994).  The star
formation history in the thin disk in the solar neighborhood is also
consistent with a fairly constant rate, only a few times higher in the
past than at present.  Ongoing star-formation rates in disks are, in
general, higher in the inner parts than in the outer disk (e.g.
Kennicutt 1989).  The outer regions of disks are clearly less-evolved
than the inner regions, having a higher gas mass fraction and a lower
mean metallicity.  Detailed analyses of ages and elemental abundances
for stars in the disk of the Milky Way are consistent with a higher
star formation rate in the early stages in the inner regions of the
disk (Edvardsson {\it et al.} 1993; Wyse 1995).  Further evidence for
age gradients in stellar disks comes from an analysis of recently-established
broad-band color gradients, taking into
account the effects of dust and metallicity variations (de Jong 1995).
Additionally,  disk scale lengths in H-$\alpha$,
tracing on-going massive star formation, are significantly larger than
the scalelengths in the V or I bands (Ryder and Dopita 1994).  Indeed, many
disk galaxies, including the Milky Way (de Geus {\it et al.} 1993) are
presently forming massive stars beyond the canonical optical edge of
the disk (A.~Ferguson, Ph.~D., in prep.).  Many theories of disk galaxy
formation and evolution predict that disks form from the `inside-out'
(e.g. Larson 1974), and that star formation rates should be a smooth
function of galactocentric radius (e.g. Wyse and Silk 1989).

It would appear that the future of the thin disk is to get larger,
growing outwards in radius.  Note that this is also expected if and
when the Large Magellanic Cloud merges with the thin disk, due to
angular momentum re-arrangement, together with some heating
in the plane.

\ls\ls
\ni\ub{CONCLUSIONS}

\ls
\ni
The age distribution, and chemical elemental abundances, of stars in
Milky Way provide constraints on theories of galaxy
formation.
The luminosity-weighted metallicity distribution of the present
retinue of dSph galaxies is dominated by the most metal-rich systems,
with [Fe/H] $\simgt -1$ dex.  This contrasts strongly with the field
halo. The more metal-rich central bulge of the Galaxy also could not
have formed by accretion of systems with similar stellar populations
to these metal-rich dwarfs, nor similar to the Magellanic Clouds.
However, accretion is on-going, as evidenced by the Sagittarius dwarf,
and the younger stellar population of each of the halo and bulge is
likely to increase.  The future of the thick disk is to get more
massive, as continuing interactions of the Milky Way with more massive
satellites (Magellanic Clouds) lead to merging.  The thin disk of the
Milky Way, and indeed of disk galaxies in general, appears to be
increasing in scalelength, and its future is to get larger.

\ls\ls
\ni\ub {ACKNOWLEDGEMENTS}
\ls
I thank the  Seaver Foundation for support.  The Center for
Particle Astrophysics is funded by the NSF.

\ls\ni\ub {REFERENCES}
\ls
\parindent=0pt

\parindent=0pt

\pp  Arnold, R. \&   Gilmore, G. 1992, MNRAS, 257, 225

\pp  Armandroff, T.E., DaCosta, G.S, Caldwell, N. \& Seitzer, P. 1993, AJ,
106, 986

\pp  Azzopardi, M. \& Lequeux, J. 1992, in {\it The Stellar Populations of
Galaxies}, eds B.~Barbury \& A.~Renzini (Kluwer, Dordrecht) p201

\apjref Binney, J. {\it et al.} 1991;\mnras;252;210
\apjref Blitz, L. \& Spergel, D. 1991;\apj;379;631

\pp  Bond, H.E. \& MacConnell, D.J. 1971, ApJ, 165, 51

\apjref Carlberg, R.G. 1990;\apj;350;505
\pp Carlberg, R.G., 1995, in proc Ringberg meeting, Oct 1994

\pp Carney, B.W. 1990, in {\it Bulges of Galaxies}, eds B.~Jarvis
\& D.~Terndrup (ESO : Garching) p26

\pp  Carney B.W., Latham D.W., Laird J.B., Aguilar L.A., 1994,
AJ, 107, 2240

\apjref Combes, F., Debbash, F., Friedl, D. \& Pfenniger, D.  1990;\aap;223;82

\pp  Conlon, E.S. {\it et al.} 1992, ApJ, 400, 273

\apjref de Geus, E., Vogel, S., Digel, S. \& Gruendl, R., 1994;ApJ;413;97

\pp de Jong, R. 1995, PhD Thesis, University of Groningen

\apjref Edvardsson, B., {\it et al.}  1993;\aap;275;101

\apjref Eggen, O., Lynden-Bell, D. \& Sandage, A. 1962;\apj;136;748

\pp Freeman, K.C. 1991, in {\it Dynamics of Disc Galaxies}, ed B.~Sundelius
(G\"oteborgs University, G\"oteborg) p15

\apjref Friel, E. 1987;\aj;93;1388

\pp Gerhard, O. \& Binney, J. 1993, in {\it Galactic Bulges}, eds
H.~Dejonghe \& H.~Habing (Kluwer, Dordrecht) p275

\apjref Gilmore, G. \& Wyse, R.F.G. 1985;\aj;90;2015

\pp Gilmore, G. \& Wyse, R.F.G. 1987, in {\it The Galaxy}, eds G.~Gilmore
\& B.~Carswell, (Reidel, Dordrecht) p247

\pp  Gilmore, G., Wyse, R.F.G. \& Jones, J.B. 1995, AJ 109, 1075

\pp  Green, E.M., Demarque P. \& King C.R. 1987,
 {\it The Revised Yale Isochrones}, (Yale University Observatory)

\apjref Hartwick, F.D.A. 1976;\apj;209;418
\apjref Holtzman, J.A. {\it et al.} 1993;\aj;106;1826
\pp Ibata, R., Gilmore, G. \& Irwin, M. 1994, Nature 370, 194

\pp  Ibata, R. \& Gilmore G., 1995, MNRAS in press

\apjref Jones, B., Klemola, A. \& Lin, D. 1994;\aj;107;1333

\apjref Kennicutt, R.C. 1989;\apj;344;689

\apjref Kennicutt, R.C., Tamblyn, P \& Congdon, C. 1994;\apj;435;22

\apjref Kroupa, P., Roser, S. \& Bastein, U. 1994;\mnras;266;412

\apjref Kruit, P.C. van der \& Searle, L. 1982;\aap;110;79
\pp Lacey, C. 1991, in {\it Dynamics of Disc Galaxies},
ed.~B.~Sundelius (G\"oteborgs University, G\"oteborg), p257

\apjref Lacey, C.G. \& Cole, S. 1993;\mnras;262;627
\pp Lacey, C.G. \& Cole, S. 1995, preprint

\pp  Laird, J.B., Rupen M.P., Carney B.W. \& Latham D.W. 1988, AJ,
 96, 1908

\apjref Larson, R.B. 1974;\mnras;169;229
\pp  Lee, M.G. {\it et al.}  1993, AJ, 106, 1420

\apjref Lewis, J.R. \& Freeman, K.C. 1989;\aj;97;139
\apjref Lin, D.N.C., Jones, B. \& Klemola, A.  1995;\apj;439;652

\pp Lizst, H. 1995, in {\it Unsolved Problems of the Milky Way},
ed.~L.~Blitz (Kluwer, Dordrecht) in press

\pp  Majewski, S. 1993, in
{\it Galaxy Evolution: The Milky Way Perspective}, ed.~S.~Majewski,
(ASP, San Francisco) p5

\apjref McWilliam, A. 1990;ApJS;74;1075
\pp  McWilliam, A., Rich, M., 1994, ApJS, 91, 749

\apjref Morrison, H., Boroson, T., \& Harding, P. 1994;\aj;108;1191

\apjref Morrison, H., Flynn, C. \& Freeman, K.C. 1990;\aj;100;1191

\pp  Nissen, P.E., Gustafsson, B., Edvardsson, B. \& Gilmore G. 1994,
 A\&A, 285, 440

\pp Ostriker, J.P. 1990 in {\it Evolution of the Universe of Galaxies} ed
R.~Kron (ASP, San Francisco) p25

\pp  Preston G.W., Beers T.C., Shectman S.A., 1994, AJ, 108, 538

\apjref Quinn, P.J., Hernquist, L. \& Fullagher, D.P. 1993;\apj;403;74

\apjref Raha. A, Sellwood, J., James, R. \& Kahn, F. 1991;Nature;352;411

\pp  Ryan, S.G. \& Norris, J.E. 1993, in
{\it Galaxy Evolution: The Milky Way Perspective}, ed.~S.~Majewski,
(ASP, San Francisco) p103

\apjref Ryder, S. \& Dopita, M. 1994;\apj;430;142

\pp  Sandage, A.R., 1969, ApJ, 158, 1115

\pp  Schuster, W. \& Nissen, P. 1989, A\&A, 222, 89

\apjref Searle, L. \& Zinn, R. 1978;\apj;225;357
\apjref Shaw, M.A. \& Gilmore, G. 1990;\mnras;242;59

\apjref Silk, J. \& Wyse, R.F.G. 1993;Phys Rep;231;293

\pp  Silk, J., Wyse, R.F.G. \& Shields, G.A. 1987, ApJ, 322, L59

\pp  Smecker-Hane, T., Stetson, P., Hesser, J. \& Lehnert M.~1994, AJ, 108,
507

\apjref Soubiran, C. 1993;\aap;274;181
\apjref Tremaine, S., Ostriker, J.P. \& Spitzer, L. 1975;\apj;196;407

\apjref T\'oth, G. \& Ostriker, J.P. 1992;\apj;389;5

\pp Unavane, M., Wyse, R.F.G. \& Gilmore, G. 1995, MNRAS, in press

\pp Wyse, R.F.G., 1995, in {\it Stellar Populations}, eds G.~Gilmore and
P.C.~van der Kruit (Kluwer, Dordrecht) in press

\apjref Wyse, R.F.G. \& Gilmore, G. 1988;\aj;95;1404

\apjref Wyse, R.F.G. \& Gilmore, G.  1992;\aj;104;144
\apjref Wyse, R.F.G. \& Silk, J. 1989;\apj;339;700

\vfill\eject
{\it Figure 1}: The distribution of B$-$V color and
[Fe/H] for metal-poor stars from the  sample of Carney {\it et al.} (1994).
Uncertainties have been ignored in the interests of clarity, and are of
order 0.1 dex in [Fe/H] and 0.05 in B$-$V. Superposed are turn-off
isochrones ({\it Revised Yale Isochrones}; Green {\it et al.} 1987) with
ages given in Gyr (8, 10, 15, 16, 17). The three metallicity ranges
delineated by the dashed lines are discussed separately.

\ls\ni
{\it Figure 2:} Metallicity distribution of seven of the dSph
companions to the Milky Way (solid histogram) compared with the field
stellar halo (dashed histogram). Only those dSph with mean metallicity
derived from spectroscopic estimates from a reasonably large sample of
stars have been included. The dSph are represented by their mean
metallicity, but one should bear in mind that an internal dispersion
of 0.2 -- 0.3 dex is typical. The upper panel shows the unweighted
distribution, the lower panel shows the dSph weighted by their
individual luminosities. The mean metallicities of the Magellanic
Clouds are indicated by arrows.

\ls
\ni
{\it Figure 3}: Iron abundance versus redshift of formation for stars from
the Evardsson {\it et al.} sample with kinematics consistent with their
having been formed in the Galactocentric range of 7--9 kpc.  Uncertainties
in [Fe/H] and in log age are $\sim 0.1$ dex.  Stars with ages corresponding
to formation redshifts $\simgt 1.5$ exist, but have been suppressed here.

\bye